# Fast and simple Complex and slow

**Moshe Schwartz**

Raymond and Beverly Sackler Faculty of Exact Sciences
School of Physics and Astronomy, Tel Aviv University,
Ramat Aviv, Tel Aviv 69978, Israel

The decay of a general time dependent structure factors is considered. The dynamics is that of stochastic field equations of the Langevin type, where the systematic generalized force is a functional derivative of some classical field Hamiltonian with respect to the field. Equations of this type are generic and describe many physical systems. It is usually believed that simple linear systems exhibit exponential decay in time ,while non linear complex systems ,such as ferromagnets at their critical point, decay slowly as a power law or stretched exponential. A necessary condition for slow decay is that the eigenvalues of the appropriate Fokker- Planck operator accumulate at zero for each momentum **q** . I argue here that when the necessary condition is obeyed, slow or fast (exponential) decay are both possible for simple linear systems as well as for complex non liner ones. The actual form of decay is determined by what structure factor we prefer to observe. An explicit family of linear models in which the "natural" structure factors decay exponentially is constructed. It is shown how a more complex structure factor decays in those models with a slow decay form. It is further shown how in general non linear systems it is always possible to find structure factors that decay exponentially. The question of actual observation of such structure factors in experiment and numerical simulations is discussed.

Stretched exponential decay is usually associated with complex systems such as glasses [1-3], polymers solutions [4,5] dielectric and viscoeleastic materials [6,7]. It has been recently suggested [8-11], however, that such a form of decay is much wider spread in condensed matter physics. In fact, it has been claimed that it is present in many non-linear systems such as the KPZ dynamical system, ferromagnets at their critical temperature etc. These systems have in common the property, that a disturbance of wave vector **q**, decays with a characteristic rate, $\omega_q$, that is proportional to $q^z$ with z>1. This may still suggest that only complex non-linear systems may show decay that is slower than exponential. Further, it may also suggest as already implicitly implied by the above, that a system can be characterized as fast or slow according to its nature of decay, namely, whether it decays exponentially or slower.

The purpose of the present article is to demonstrate that the above suggestions are not necessarily correct. It will be shown that the nature of decay is strongly influenced by the time dependent correlations observed to such an extent that a complex system showing slow decay for certain correlation will show fast, exponential decay for others, while simple systems that yield exponential decay for some correlations exhibit slow decay in other correlations. It is true that the experimental observation of certain correlations, like density correlations in liquids or spin correlations in ferromagnets for which there are standard measuring probes, is much more natural than is the observation of other correlations, so that in that sense it is meaningful to speak about fast and slow decaying systems. It has to be noted, however, that in numerical simulations it is relatively easy to observe any arbitrary time dependent correlation.

Consider first the general case of a field evolving according to a Langevin equation in the Fourier components of the field, $\varphi_\mathbf{q}$,

$$\frac{d\varphi_\mathbf{q}}{dt} = -\gamma \frac{\partial W}{\partial \varphi_{-\mathbf{q}}} + \eta_\mathbf{q}, \qquad (1)$$

where $W$ is the energy functional and the noise, $\eta_\mathbf{q}$ obeys

$$<\eta_\mathbf{q}(t)> \quad \text{and} \quad <\eta_\mathbf{q}(t)\eta_{-\mathbf{q}}(t')> = 2\gamma kT\delta(t-t'). \qquad (2)$$

The Langevin equation can be replaced in a standard manner by the corresponding Fokker-Planck equation for the probability distribution, to find a given field configuration,

$$\frac{\partial P}{\partial t} = \sum \gamma \frac{\partial}{\partial \varphi_{\mathbf{q}}} [kT \frac{\partial}{\partial \varphi_{-\mathbf{q}}} + \frac{\partial W}{\partial \varphi_{-\mathbf{q}}}] P . \qquad (3)$$

This has the Gibbs distribution $P_{eq} \propto \exp[-W/kT]$ as an equilibrium solution and a standard similarity transformation induced by

$$P = P_{eq}^{1/2} \psi , \qquad (4)$$

leads to an equation for $\psi$, that looks like an imaginary time Schrödinger equation,

$$\frac{\partial \psi}{\partial t} = - \mathcal{H} \psi , \qquad (5)$$

where $\mathcal{H}$ is Hermitian, with all eigenvalues positive excluding zero that corresponds to the ground state. The Hamiltonian $\mathcal{H}$ is given by

$$\mathcal{H} = -\gamma kT \sum [\frac{\partial}{\partial \varphi_{\mathbf{q}}} - 1/(2\gamma kT)\frac{\partial W}{\partial \varphi_{\mathbf{q}}}][\frac{\partial}{\partial \varphi_{-\mathbf{q}}} + 1/(2\gamma kT)\frac{\partial W}{\partial \varphi_{-\mathbf{q}}}] . \qquad (6)$$

Let $[\chi]_q$ be the Fourier transform of some general composite field $\chi$. Consider the time dependent correlation function, $S_{\chi,\mathbf{q}} = <[\chi(0)]_{-\mathbf{q}}[\chi^*(t)]_{\mathbf{q}}>$, measured at equilibrium. The meaning of the above is that when the system is at equilibrium $[\chi]_{-\mathbf{q}}$ is measured at time t=0 the system is allowed then to evolve freely and $[\chi^*]_{\mathbf{q}}$ is measured at a later time, t. It is easy to show [8] that

$$S_{\chi,\mathbf{q}}(t) = <0|[\chi]_{-\mathbf{q}} \exp[-\mathcal{H}t][\chi^*]_{\mathbf{q}}|0> , \qquad (7)$$

where $|0\rangle$ is the ground state of $\mathcal{H}$, that in the coordinate ($\varphi$) representation is $P_{eq}^{1/2}$.

Since the state $[\chi^*]_\mathbf{q}|0\rangle$, carries a definite momentum $\mathbf{q}$, the correlation function can be written in a more transparent way,

$$S_{\chi,\mathbf{q}}(t) = \sum_\beta |<n;\mathbf{q},\beta|[\chi]_\mathbf{q}|0>|^2 \exp[-\lambda_{\mathbf{q},\beta}t] \qquad (8)$$

where $|n;\mathbf{q},\beta>$ denotes an eigenstate of $\mathcal{H}$ carrying momentum $\mathbf{q}$, with $\beta$ standing for all the other "quantum" numbers and $\lambda_{\mathbf{q},\beta}$ is the corresponding eigenvalue of $\mathcal{H}$.

It is clear that two necessary conditions have to be met in order to have a decay that is slower than exponential. The first is that the set $\{\lambda_{\mathbf{q},\beta}\}$ accumulates at zero for fixed $\mathbf{q}$. (Indeed, it has been recently shown [12], that when the parameters of $W$ correspond to a system at a second order phase transition, the eigenvalues of the FP operator, corresponding to a given $\mathbf{q}$, accumulate at zero as the size of the system tends to infinity). The second requirement is that the matrix elements do not vanish too strongly as the corresponding eigenvalues tend to zero.

A soluble linear example is considered in the following. It is shown that time dependent structure factors of polynomials in the basic field decay exponentially fast. An explicit example is then given of the structure factor of some composite field that exhibits slow decay. One of the basic propositions of this article is thus demonstrated.

Consider the bilinear energy functional

$$W = \frac{A}{2}\sum q^z \varphi_\mathbf{q}\varphi_{-\mathbf{q}}, \qquad (9)$$

in $d$ dimensions with $z>2$ (This is a generalization of the Edwards-Wilkinson system [10], for which $z=2$) The corresponding "Hamiltonian" is

$$\mathcal{H} = A\sum q^z \xi_\mathbf{q}^* \xi_\mathbf{q}, \tag{10}$$

where the $\xi$'s obey the usual Bose commutation relations.

$$[\xi_\mathbf{q},\xi_\mathbf{p}] = [\xi_\mathbf{q}^*,\xi_\mathbf{p}^*] = 0 \text{ and } [\xi_\mathbf{q},\xi_\mathbf{p}^*] = \delta_{\mathbf{q},\mathbf{p}} \tag{11}$$

and are related to the field and field derivative by

$$\varphi_\mathbf{q} = [\gamma kT / Aq^z]^{1/2}[\xi_\mathbf{q} + \xi_{-\mathbf{q}}^*] \text{ and } \frac{\partial}{\partial \varphi_\mathbf{q}} = [Aq^z / 4\gamma kT]^{1/2}[\xi_{-\mathbf{q}} - \xi_\mathbf{q}^*] \tag{12}$$

It is easy to verify that the eigenvalues of $\mathcal{H}$ in this simple system accumulate at zero. This fact is easiest seen by considering an eigenstate of $\mathcal{H}$ of total momentum $\mathbf{q}$ that has $n$ excitations each carrying momentum $\mathbf{q}/n$. The corresponding eigenvalue is

$$\lambda_{\mathbf{q},n} = nA[q/n]^z = Aq^z n^{1-z} \tag{13}$$

since $n$ can be made arbitrarily large it is obvious that $\{\lambda_{\mathbf{q},n}\}$ tends to zero for any given $\mathbf{q}$ as $n$ tends to infinity. In spite of that, the "natural" time dependent structure factor is easily calculated to yield:

$$<\varphi_{-\mathbf{q}}(0)\varphi_\mathbf{q}(t)> = [\gamma kT / Aq^z]\exp(-Aq^z t) \; . \tag{14}$$

The reason for the simple exponential decay is lack of matrix elements. The only matrix elements of the form $<n;\mathbf{q}|\varphi_\mathbf{q}|0>$ that do not vanish are those for which $|n;\mathbf{q}> = \xi_\mathbf{q}^*|0>$. Taking the composite field $\chi(\mathbf{r})$ to be a polynomial in the field $\varphi(\mathbf{r})$, it still follows from eq. (8) that the corresponding structure factor decays faster than some exponential in time, since all the matrix elements of the form $<n;\mathbf{q}|\chi_\mathbf{q}|0>$ vanish for states with a number of excitations, that exceeds the degree of the polynomial. It is thus clear that in order to have a decay slower than exponential, it is necessary that $\chi(\mathbf{r})$ is a composite field of infinite order in $\varphi$. Indeed, it is easy to construct such an example. Take the composite field $\chi(\mathbf{r}) = \exp[i\alpha\varphi(\mathbf{r})]$, where $\alpha$ is a constant that has the dimensions of $\varphi^{-1}$. It is straightforward to show that the corresponding time dependent structure factor is

$$S_{\chi,\mathbf{q}}(t) = \int d^d\mathbf{r}\exp[-(\alpha^2/2)w(\mathbf{r},t) - i\mathbf{q}\cdot\mathbf{r}] \quad , \tag{15}$$

where

$$w(\mathbf{r},t) = [4\gamma kT / A^{d/z}]f(r/(At)^{1/z}). \tag{16}$$

The function $f$ is defined by

$$f(x) = (1/2\pi)^{d+1}\int d^d\mathbf{u}d\omega[1 - \cos[(\mathbf{u}\cdot\mathbf{x}) + \omega]]/[\omega^2 + u^{2z}]. \tag{17}$$

It is simplest to study the case z>d. In that case an upper cut-off is not needed to make $f$ finite. Focus now on the asymptotic form of $f$ for small $q$ ($q << [\gamma kT\alpha^z / A]^{1/(2d-z)}$) and long times ($Aq^z t >> 1$).

The asymptotic behavior of $S_{\chi,q}(t)$ in the above regime is determined by contributions to the integral on the right hand side of eq. (15) from regions where $r/(At)^{1/z}$ is small. Therefore,

$$S_{\chi,\mathbf{q}}(t) = [\pi/\gamma kT\alpha^2 tf''(0)]^{d/2}(At)^{d/z(d/2+1)} * \tag{18}$$

$$* \exp\{-[2\gamma kT\alpha^2 f(0)t/(At)^{d/z} + q^2(At)^{(d+2)/z}/(4\gamma kT\alpha^2 f''(0)t)]\}$$

Since z>d and z $\geq 2$ it is clear that the exponential is stretched in time. (Note that both $f(0)$ and $f''(0)$ are finite positive numbers). The structure factor of the composite field is thus shown to decay slower than an exponential.

The opposite direction, namely the proof that in non-linear systems, for which the eigenvalues of the FP operator accumulate at zero for each $\mathbf{q}$ it is always possible to find structure factors that decay exponentially in time, is straightforward. Choose $\chi_{\mathbf{q}}$, such that $|\psi_{\mathbf{q}}\rangle = \chi_{\mathbf{q}}|0\rangle$ is an exact eigenfunction of $\mathcal{H}$. A simple example for $\chi_q$ is to obtain some $|\psi_q\rangle$ in the coordinate representation, $\psi_{\mathbf{q}}\{\varphi\}$ and write $\chi_{\mathbf{q}}\{\varphi\} = \psi_{\mathbf{q}}\{\varphi\}/P_{eq}^{1/2}\{\varphi\}$. Obtaining an exact excited state of $\mathcal{H}$ is usually impossible. So although, in principle, it is possible to find a simple mode of the system that decays exponentially, practically this is quite difficult. In fact, one can only expect to obtain successive approximations for $\psi_q$, resulting in a decay that may look exponential for times that become longer with the order of approximation. We see that, in principle, we can find for such a system correlations that decay exponentially but is it clear that we can always obtain slow decay? The positive answer is almost trivial. It is possible to find an infinite set of eigenstates of the "Hamiltonian", $\{|\psi_{\mathbf{q}}^\beta\rangle\}$, that carry a definite momentum $\mathbf{q}$ and are perpendicular to each other. The corresponding eigenvalues $\{\lambda_{\mathbf{q}}^\beta\}$ accumulate at 0. This means that a partial sequence of the eigenvalues, $\{\lambda_{\mathbf{q}}^i\}$, can be found that tends to zero as $i$ tends to infinity. Choose $\chi_{\mathbf{q}}^i$, such that $|\psi_{\mathbf{q}}^i\rangle = \chi_{\mathbf{q}}^i|0\rangle$. This can be done exactly as described above. Construct now

$$\chi_{\mathbf{q}}^* = \sum_{i=1}^{\infty} A_i^*(\mathbf{q})\chi_{\mathbf{q}}^{i*}, \qquad (19)$$

such that $\sum_{i=1}^{\infty}|A_i(\mathbf{q})^2|$ is finite. It is clear that the $\chi$ structure factor

$$S_{\chi\mathbf{q}}(t) = \sum_{i=1}^{\infty}|A_i(\mathbf{q})|^2 \exp[-\lambda_{\mathbf{q}}^i t] \qquad (20)$$

can be made to have any desired form of long time decay by choosing the coefficients $A_i(\mathbf{q})$ accordingly. In particular, any form of slow decay such as stretched exponential, power law etc. can be obtained. In the following it will become clear why slow decay is the rule rather than the exception.

The conclusion of the above is the following. Any system with eigenvalues of the FP operator that accumulate at zero for each **q**, whether linear or non-linear, has correlations of "simple" quantities that decay exponentially in time while the correlations of composite quantities may decay much slower in time. It is clear that the relative measure of quantities with exponentially decaying correlations is zero. (To see this think what are the limitations on the coefficients in eq. (19) above, such that the decay is still exponential. The relative measure is actually the relative measure of decaying exponentials to all other slower decaying functions.) Therefore a random choice of a quantity to be measured will always produce slow decay. The reason why in linear systems we usually obtain exponential decay is because in those systems our choice is not random. The "simple" quantities are also the quantities we would naturally be interested in, like $\varphi_q$, the Fourier transform of the field. Thus, slow or fast form of decay is not a property of the system it is the property of the specific correlations we are measuring. "Simple" correlations will decay fast and composite correlations will decay slowly. A practical question is whether "natural" correlations are also simple.